\documentclass[twocolumn,aps,prb]{revtex4}
\usepackage{graphicx}
\begin{document}
%
\title{Critical behavior of
magnetic systems with extended impurities in general dimensions}
\author{V.~Blavats'ka}
\email[]{viktoria@icmp.lviv.ua}
\affiliation{Institute for Condensed Matter
  Physics of the National Academy of Sciences of Ukraine, 79011 Lviv,
  Ukraine}
\affiliation{Theoretical Polymer Physics, Hermann-Herder-Str. 3, 79104
  Freiburg University, Germany}
\author{C.~von~Ferber}
\email[]{ferber@physik.uni-freiburg.de}
\affiliation{Theoretical Polymer Physics, Hermann-Herder-Str. 3, 79104
  Freiburg University, Germany}
\author{Yu.~Holovatch}
\email[]{hol@icmp.lviv.ua}
\affiliation{Institute for Condensed Matter
  Physics of the National Academy of Sciences of Ukraine, 79011 Lviv,
  Ukraine}
\affiliation{Ivan Franko National University of Lviv, 79005
  Lviv, Ukraine}
\begin{abstract}
 We investigate the critical properties of $d$-dimensional
 magnetic systems with quenched extended defects, correlated in
 $\varepsilon_d$ dimensions (which can be considered as the
 dimensionality of the defects) and randomly distributed in the
 remaining $d-\varepsilon_d$ dimensions; both in the case of fixed
 dimension $d=3$ and when the space dimension continuously changes
 from the lower critical dimension to the upper one. The
 renormalization group  calculations are performed in the minimal
 subtraction scheme.  We analyze the two-loop
 renormalization group functions for different
 fixed values of the parameters $d, \varepsilon_d$.  To this end, we
 apply the Chisholm-Borel resummation technique and report the
 numerical values of the critical exponents for the universality class
 of this system.
\end{abstract}
\date{\today}
\maketitle
\section{Introduction}

The study of critical properties of the various kinds of disordered
systems remains one of the central problems in condensed matter
physics.  We focus our attention here to the case of $d$-dimensional
magnetic systems with quenched nonmagnetic impurities.  The first
question that arises here is: how does quenched disorder influence
criticality, i.e., will the universal scaling laws, governing the
critical behavior of the ``pure" system be altered by quenched
disorder introduced into the system in a quenched manner?

The effect of weak quenched uncorrelated point-like disorder on the critical
behavior of magnetic systems is predicted by the Harris criterion
\cite{Harris74}:
disorder changes the critical exponents
only if the critical exponent $\alpha_{p}$ of the
pure (undiluted) system is positive:
\begin{equation}
\alpha_p=2-d\nu_p>0,
\label{harris}
\end{equation}
$\nu_p$ being the correlation length critical exponent of the pure
system. Of the $d$-dimensional spin systems that are described by the
$m$-vector model, only the pure Ising model is characterized by a value of
$\alpha_{p}>0$ and thus is affected by point-like weak disorder at
criticality.

Real magnetic crystals often contain defects in the form of linear
dislocations, planar grain boundaries, 3-dimensional cavities or
regions of different phases, embedded in the matrix of the original
crystal, as well as various complexes (clusters) of point-like
non-magnetic impurities \cite{defectbook}.  Systems with such
``extended'' (macroscopic) defects have attracted much interest
\cite{Dorogovtsev80,Boyanovsky82,Lawrie84,Yamazaki86,Yamazaki,%
Weinrib83,Ballesteros99,Prudnikov99,Lee92,Cesare94,%
Korzhenevskii96,Blavatska}.

Dorogovtsev \cite{Dorogovtsev80} proposed the model of a
$d$-dimensional $m$-component spin system with quenched random
nonmagnetic impurities, that are strongly correlated in
$\varepsilon_d$ dimensions and randomly distributed over the remaining
$d-\varepsilon_d$ dimensions.  Such a system is no longer isotropic;
the idea of two different correlation lengths naturally arises since
the system is expected to behave differently along the directions
``parallel'' to the $\varepsilon_d$-dimensional impurity and along the
``perpendicular'' directions. The case $\varepsilon_d=0$ is associated
with point-like defects, and extended parallel linear (planar) defects
are related to the cases $\varepsilon_d$ = 1(2).
Generalizing $\varepsilon_d$ to non-negative real numbers
it may be interpreted as
an effective fractal dimension of a complex random defect system \cite{Yamazaki86}.
The critical behavior of such a model was examined by means of the
renormalization group (RG) method
\cite{Dorogovtsev80,Boyanovsky82,Lawrie84}.
A double expansion in
both $\varepsilon=4-d$, $\varepsilon_d$ was suggested and
RG functions were calculated \cite{Dorogovtsev80} to
order $\varepsilon$, $\varepsilon_d$ ; qualitatively,
the crossover to a new universality class in the
presence of extended defects was found. These calculations were
extended to the second order in Ref.\cite{Boyanovsky82}. Here, it was
argued, that the Harris criterion is modified in the presence of
extended impurities: the randomness is relevant, if
\begin{equation}\label{criterion}
\varepsilon_d>d-\frac{2}{\nu_{p}}. \label{genharris}
\end{equation}
For point-like disorder $\varepsilon_d=0$ the Harris
criterion (\ref{harris}) is restored from (\ref{genharris}). In
particular, (\ref{genharris}) defines for each value
of $m$ a lower marginal defect dimension
$\varepsilon_d^{{\rm marg}}$, above which the critical exponents
are influenced by disorder. Note that for a negative value of
$\varepsilon_d^{{\rm marg}}$
any amount of quenched impurities will induce new
critical behavior as far as $\varepsilon_d$ is always
positive.

\begin{figure}[htbp] \begin{center}
\includegraphics[width=85mm]{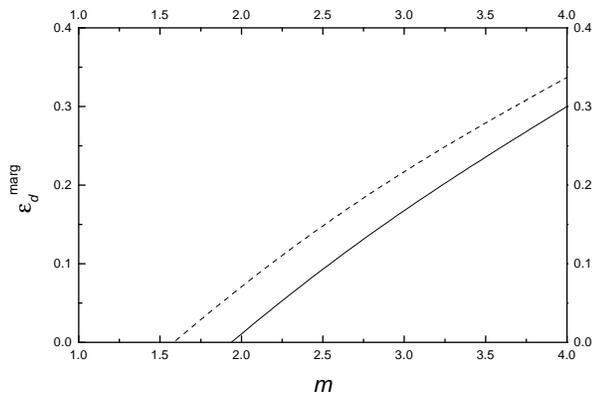} \end{center}
\caption{\label{fig1}
The marginal value of $\varepsilon_d^{\rm marg}$
for  3-dimensional
$m$-component magnetic systems as function of
$m$. Disorder is relevant in the region  of $\varepsilon_d, m$ plane
above the $\varepsilon_d^{\rm marg}$ curve.  The solid line is obtained
by substituting
into \protect (\ref{criterion}) the 6-loop results
\protect \cite{Guida98} for the critical exponents of pure $m$-vector
magnet.
The dashed line shows our present results, obtained in 2-loop
approximation.}
\end{figure}

Taking the best known estimates from a 6-loop RG expansion for the
exponent $\nu_p$ of the different $m$-component
systems\cite{Guida98}, one finds that the disorder with extended
defects is relevant for $d=3$ over a wider range of $m$ than the
point defect disorder. We show the lower marginal value
$\varepsilon_d^{\rm marg}$ for these systems in Fig. \ref{fig1}.
Its asymptotic value for large $m$ is $\varepsilon_d^{{\rm
marg}}(m=\infty)=1$. This estimate is easily obtained
using the value of the correlation length critical exponent
for $d=3$ spherical model \cite{Kac52} $\nu(d=3, m=\infty)=1$ in
(\ref{genharris}).

Although the RG functions of Dorogovtsev's model \cite{Dorogovtsev80}
were obtained  in Ref. \cite{Boyanovsky82} within the two loop
accuracy, a numerical analysis was provided only to order
$\varepsilon$, $\varepsilon_d$. For the class of extended-defect
systems with cubic anisotropy, the second order
$\varepsilon,\varepsilon_d$ RG analysis has been performed
\cite{Yamazaki}.

A related model with long-range-correlated quenched disorder, has been
proposed in the work of Weinrib and Halperin \cite{Weinrib83} and
was recently studied in Refs. \cite{Prudnikov99,Blavatska}.  This
model is characterized by a correlation function that has a power law
decay $g(r)\sim r^{-b}$ with distance $r$.  This type of disorder has
a direct interpretation for integer values of $b$.  Namely, the case
$b=d$ corresponds to point-like defects, while $b=d-1(b=d-2)$
describes straight lines (planes) of impurities of {\it random}
orientation, in contrast to the {\it parallel} extended impurities,
considered above.  The results of Weinrib and Halperin were confirmed
by a Monte Carlo study\cite{Ballesteros99} of the 3-dimensional random
Ising model with linear defects of random orientation.  An extension
of this model describes $d$-dimensional systems with quenched disorder
that is completely correlated in $\varepsilon_d$ dimensions and
long-range correlated in the remaining $d-\varepsilon_d$ dimensions
\cite{Cesare94},\cite{Korzhenevskii96}. This model applies when
parallel dislocations in a crystal induce an elastic stress field in
the direction perpendicular to the dislocation axis. The stress field
may couple to the local critical temperature and play the role of a
random potential. Real linear dislocations are described by
$\varepsilon_d=1$. The case $\varepsilon_d=0$ reproduces the model of
Weinrib and Halperin. The renormalization group treatment has been
performed in a one-loop expansion within triple
($\varepsilon,\varepsilon_d, \sigma=d-b$) \cite{Cesare94} and
($\varepsilon,\varepsilon_d,\delta=4-b$) \cite{Korzhenevskii96}
expansions.

Although systems with extended quenched defects are subject to a
number of studies, a complete classification of the critical behavior
of magnets of this type has yet to be established. The results that
have been obtained so far using second order double expansions are of
qualitative character only.  The series of the RG functions are known
to be asymptotic at best, and to obtain reliable quantitative results,
appropriate resummation techniques should be applied.  Such have
proved to be fruitful for the analysis of point-like quenched disorder
\cite{Jug83,reviews}. In our previous paper \cite{slovakia} we have
already applied this analysis to the 3-dimensional magnetic systems
with extended defects.  Previously, resummation techniques were not
exploited for these systems, apart from the special case $d=3$ and
$\epsilon_d=1$, corresponding to linear defects \cite{Lawrie84}.
There, critical exponents were calculated to second order in an
expansion in $\varepsilon$, $\varepsilon_d$ and Pad\'e-like
approximants were used to provide numerical estimations.  In our
present paper we explore a wider region of the phase diagram shown in
Fig. \ref{fig1} and apply resummation schemes to estimate
numerically the critical exponents of extended-defect systems both in
the case of space dimension $d=3$ and in general dimensions,
using the two loop RG functions of Refs. \cite{Boyanovsky82,Lawrie84}.

Our paper is organized as follows. In the following section II we
present the model.  The field-theoretical analysis of critical
behavior is reviewed in section III, and expressions are given for the
RG functions.  Resummation is applied to the RG
functions both at $d=3$ and in the case of general dimensions and
numerical estimates for the critical exponents are presented in
section IV followed by conclusions in section V.

\section{The model}

We consider the model \cite{Dorogovtsev80,Boyanovsky82,Lawrie84}
of an $m$-vector magnet with $\varepsilon_d$-dimensional
impurities, each extending throughout the system along the
coordinate directions symbolized as $x_{||}$, whereas in the
remaining $d-\varepsilon_d$ dimensions they are randomly
distributed. The effective Hamiltonian of the model reads
\cite{Boyanovsky82,Lawrie84}:
\begin{eqnarray}
{\cal H}&=&\int
d^dx\big[\frac{1}{2}(\mu_0^2+V(x)) \vec{\phi}^2(x)+
(\nabla_{\perp}
\vec{\phi}(x))^2 \nonumber\\
&& +a_0({\nabla}_{||}
\vec{{\phi}}(x))^2+\frac{u_0}{4!}(\vec{\phi}^2(x))^2\big].
\end{eqnarray}
Here, $\vec{\phi}$ is an $m$-component vector field:
$\vec{\phi}=\{ \phi^{1}\cdots\phi^{m}\}$, $\mu_0$ and $u_0$ are
the bare mass and the coupling of the magnetic model, $a_0$ is the
bare anisotropy constant, $\nabla_{||}$ and $\nabla_{\perp}$ stand
for differentiation in the coordinates $x_{||}$ (along
$\varepsilon_d$ directions) and $x_{\perp}$ (along
$(d-\varepsilon_d)$ directions) respectively, $V(x)$ represents
the impurity potential.  The probability distribution for the
impurities is defined to yield:
\begin{eqnarray}
\langle\langle V(x)\rangle\rangle&=&0,\nonumber \\
\langle\langle V(x)V(y)\rangle\rangle&
=&-v_0\delta^{d-\varepsilon_d}(x_{\perp}-y_{\perp}). \,\,\,\,\,\,
\label{corr}
\end{eqnarray}
Here, $\langle\langle...\rangle\rangle$ stands for the average
over the potential distribution, (-$v_0$) is a positive constant
proportional to both the concentration of impurities and the
strength of their potential. The extended defects make the space
coordinate anisotropic; the anisotropy is parameterized by the constant
$a_0$.  It can initially be assigned a value of
unity, but will be renormalized away from this value in the scaling limit
\cite{Lawrie84}.

As far as the disorder is considered to be quenched, the free energy of
the system is obtained as follows \cite{Brout59}:
\begin{equation}
{\cal F}=-kT\langle\langle \ln {\cal Z}_c\rangle\rangle,
\label {replica}
\end{equation}
where ${\cal Z}_c$ is the partition function for a given (quenched)
configuration of impurities, and the average is performed over the
impurity probability distribution (\ref{corr}).  To avoid averaging
the logarithm of the partition function in (\ref{replica}), the
standard procedure is to use the replica trick \cite{Emery75}. This
amounts to formally implementing the identity $\ln{\cal Z}=\lim_{n\to
  0}({\cal Z}^n-1)/n$ by analytic continuation in $n$.  After performing
the average the $n$-replicated effective Hamiltonian then reads
\cite{Dorogovtsev80}:
\begin{eqnarray}
{\cal H}&=&\sum_{\alpha=1}^n\int{\rm d}^d x
\big\{\frac{1}{2}[\mu_0^2\vec{\phi}_{\alpha}^2(x)+
(\nabla_{\perp}\vec{\phi} _{\alpha}(x))^2 \nonumber\\
&+&a_0(\nabla_{||}\vec{\phi}_{\alpha}(x))^2]+
\frac{u_0}{4!}(\vec{\phi}_ {\alpha}^2(x))^2 \big\} \label{hamiltonian} \\
&+&\frac{v_0}{2} \sum_{\alpha,\beta=1}^n\int{\rm d}^d x\int{\rm
d}^d y
\,
\delta^{d-\varepsilon_d}(x_{\perp}-y_{\perp})\vec{\phi}_
{\alpha}^2(x)\vec{\phi}_{\beta}^2(y).\nonumber
\end{eqnarray}
Here, Greek indices denote replicas. The last term introduces an additional
coupling $v_0<0$. It is present only for non-zero dilution and is directly
responsible for the effective interaction between replicas due to the
presence of impurities. The  replica limit $n\to 0$ is implied for
all quantities calculated in the following.

The model described by the effective Hamiltonian (\ref{hamiltonian})
has a rich scaling behavior. An introduction to its critical
exponents and a derivation of their scaling relations may be found
in Refs. \cite{Dorogovtsev80,Boyanovsky82,Lawrie84}. Here, we
summarize the main quantities that describe the
critical behavior of the model (\ref{hamiltonian}). Due to the
spatial anisotropy two correlation lengths exist, one perpendicular
and one parallel to the extended impurities direction: ($\xi_{\perp}$
and $\xi_{||}$). As the critical temperature $T_c$ is approached,
their divergences are characterized by corresponding critical
exponents  $\nu_{\perp}$, $\nu_{||}$:
\begin {equation}
\label{xi} \xi_{\perp}\sim
|t|^{-\nu_{\perp}},\phantom{5555555}
\xi_{||}\sim |t|^{-\nu_{||}},
\end{equation}
where $t$ is the reduced distance to the
critical temperature $t=(T-T_c)/T_c$.  The correlation of
the order parameter fluctuations in two different points depends on
the orientation of their distance vector \cite{Dorogovtsev80}. Thus, the
critical exponents $\eta_{\perp}$ and $\eta_{||}$, that characterize the
behavior of the correlation function in the directions, perpendicular and
parallel to the extended defects, must be distinguished. On the
other hand, as far as the interaction of all order parameter
components with defects is the same, the system susceptibility is
isotropic \cite{Dorogovtsev80} and can be expressed by the pair
correlation function \cite{Yamazaki86}:
\begin{eqnarray}
\chi(k_{\perp},k_{||},t)&=&
|t|^{-\gamma}g(\frac{k_{\perp}}{|t|^{\nu_{\perp}}},
\frac{k_{||}}{|t|^{\nu_{||}}},\pm 1)\nonumber\\
&=&
\left\{
 \begin{array}{ll}
k_{\perp}^{\eta_{\perp}-2}
g(1,\frac{k_{||}}{k_{\perp}^{\nu_{||}/\nu_{\perp}}},
\frac{|t|}{k_{\perp}^{1/\nu_{\perp}}})
,& \\ k_{||}^{\eta_{||}-2}
g(\frac{k_{\perp}}{k_{||}^{\nu_{\perp}/\nu_{||}}},1,
\frac{|t|}{k_{||}^{1/\nu_{||}}})
.& \end{array}\right.
\label{chi1}
\end{eqnarray}
In (\ref{chi1}), $k_{||}$, $k_{\perp}$ are the components of the
momenta along $\varepsilon_d$ and $d-\varepsilon_d$ directions,
respectively, $\gamma$ is the magnetic susceptibility
critical exponent  and $g$ is the scaling function.
For the above introduced critical exponents the following scaling
relations hold \cite{Boyanovsky82,Lawrie84,Yamazaki86}:
\begin{equation}
\gamma=(2-\eta_{\perp})\nu_{\perp}=(2-\eta_{||})\nu_{||}.
\label{scaling1}
\end{equation}
The critical exponent $\alpha$ of the specific heat is related to
$\nu_{\perp}$, $\nu_{||}$ by another scaling relation that differs
from the ordinary one \cite{Dorogovtsev80}:
\begin{equation}
\alpha=2-(d-\varepsilon_d)\nu_{\perp}-\varepsilon_d\nu_{||}.
\label{scaling2}
\end{equation}
All the other scaling relations are of the standard form
\cite{Stanley71}.
This implies that one should calculate at least three independent exponents
(e.g., $\nu_{||},\nu_{\perp},\gamma$) instead of two, as in the standard
case, to find the others by scaling relations.


\section{Renormalization group functions}

To describe the long-distance properties of the model
(\ref{hamiltonian}) near the second order phase transition we use
the field-theoretical renormalization group (RG) method
\cite{rgbooks}. In this approach one observes the behavior of the
system under a scaling transformation that rescales the length
e.g. by a factor $\ell$. At the critical point the system assumes
scale invariance. Away from  the critical point the change of
couplings $u_0,v_0\to u,v$ and the anisotropy constant $a_0\to a$
under rescaling defines a flow in parametric space, expressed by
RG functions:
\begin{equation}
\beta_u=\frac{\partial u}{\partial \ln \ell}|_0,
\phantom{5555}\beta_v=\frac{\partial v}{\partial \ln \ell}|_0,
\phantom{5555}\zeta_{a}=\frac{\partial \ln a}{\partial \ln
\ell}|_0.
\label{13}
\end{equation}
Here, $\ell$ is the rescaling factor and the notation $|_0$ indicates
differentiation at fixed bare parameters.  The bare field $\phi$ and
the bare mass $\mu_0$ are related to the renormalized field $\varphi$
and mass $\mu$ by:
\begin{equation} \phi=Z_{\varphi}^{1/2}\varphi;
\phantom{55555}\mu_0^2=Z_{\mu^2}\mu^2.
\end{equation}
The fixed points (FPs) $u^*,v^*$ of the RG transformation are
the solutions of the system of FP equations:
$\beta_u(u^*,v^*)=0$, $\beta_{v}(u^*,v^*)=0$.
A FP is stable if the eigenvalues of the stability matrix
$
B_{ij}={\partial \beta_{u_i}(u^*,v^*)}/{\partial \beta_{u_j}(u^*,v^*)}
$
have positive real parts. The stable
FP that can be reached starting from the initial values
of the coupling constants (in our case, $u>0$, $v<0$), corresponds to
the critical point of the system. At this point, the magnetic
susceptibility and correlation length critical exponents  are given by
by the relations:
\begin{eqnarray}
\gamma^{-1}&=& 1-\frac{{\bar
\gamma}_{\phi^2}}{2-\gamma_{\phi}},\nonumber\\ \nu_{\perp}^{-1}&=&2-
{\bar \gamma}_{\phi^2}-\gamma_{\phi},\label{exp}\\
\nu_{||}&=&(1-\frac{\zeta_{a}}{2})\nu_{\perp}\nonumber,
\end{eqnarray}
while the expressions for the RG functions $\gamma_{\phi},{\bar
\gamma}_{\phi^2}$ read:
\begin {equation}
\gamma_{\phi}=\frac{\partial
\ln Z_{\varphi}}{\partial \ln \ell}|_0,\phantom{5555}
{\bar \gamma}_{\phi^2}=\frac{\partial \ln Z_{\mu^2}^{-1} }{\partial
\ln \ell }|_0-\gamma_{\phi}.
\end{equation}
The other critical exponents can be obtained from
the scaling relations (\ref{scaling1}), (\ref{scaling2}).

In order to derive the quantitative characteristics of the critical
behavior of  magnetic systems with extended impurities, we analyze the
2-loop RG functions, obtained in \cite{Boyanovsky82} in the minimal
subtraction scheme.  Unfortunately, that paper contains some numerical
errors, as pointed out in Ref. \cite{Lawrie84}. The appropriately
corrected functions read:
\begin{eqnarray}
{\beta_u}/{u}&=&-\varepsilon+\frac{(m+8)}{6}u-2v-
\frac{(3m+14)}{12}{u}^{2}
\nonumber\\
&+& \frac{1}{12}uv\left[
\frac{2}{3}(11m+58)+(m-4)\frac{\varepsilon_d}{3(\varepsilon+
\varepsilon_d)}\right]
\nonumber\\
&-&
v^2\frac{1}{144}\left[
328+32\frac{\varepsilon_d}{\varepsilon+\varepsilon_d}\right]; \nonumber\\
{\beta_{v}}/{v}&=&-\varepsilon-{\varepsilon_d}-\frac{4}{3}v+
\frac{m+2}{3}u-\frac{7}{6}
v^2\nonumber\\
&+&v u\frac{m+2}{18}\left[11-\frac{\varepsilon_d}{\varepsilon+
\varepsilon_d}\right]-\frac{5}{12}\frac{m+2}{3}u^2
; \label{beta}\\
\gamma_{\phi}&=& \frac{1}{36}v^2-\frac{m+2}{36}v u +\frac{m+2}{72}u^2
; \nonumber\\
{\bar \gamma}_{\phi^2}&=&u\frac{m+2}{6}-\frac{1}{3}v-\frac{m+2}{6}u^2-24v^2
\nonumber\\
&+&\frac{m+2}{24}v u \left[6-\frac{\varepsilon_d}{\varepsilon+
\varepsilon_d} \right];\label{gamma}\\
\zeta_{a}&=&-\frac{1}{3}v-\frac{5}{36}v^2+\frac{1}{36}(2+m)v\,u.
\nonumber
\end{eqnarray}
In the case of $\varepsilon_d=0$ the dimension enters the
minimally subtracted RG $\beta$-functions (\ref{beta}) only
in the form of the
engineering dimension $\varepsilon$. For fixed dimension
$\varepsilon=1$ the flow (\ref{13}) with the $\beta$-functions
(\ref{beta}) can then be directly
evaluated \cite{minsub,Folk00}.  Also for non-zero $\varepsilon_d$
no singularities are introduced by the dimension dependence.  We
thus propose to extend the approach of direct evaluation to the RG
functions of the present model, i.e. to treat them directly at
$d=3$ $(\varepsilon=1)$ for different fixed values of the
(fractal) defect dimensionality $\varepsilon_d$.

The expansions of the RG functions in powers of the coupling are known
for a number of models with high accuracy. However, the very idea of
perturbation theory is that the result may successfully be approximated
by accounting for higher order contributions. In the field-theoretical RG
approach the series appear to be divergent; moreover, they are
characterized by a factorial growth of the coefficients implying a
zero radius of convergence \cite{rgbooks}. A simple method to manage
this problem is to  truncate the series optimally, i.e., to account
only for those first several terms that do not show the divergence.  To take
into account the higher order contributions, the application of
special tools of resummation is required \cite{Hardy48}.

There remains the principal question about the Borel summability of
the perturbation series for a given model;
at present such a proof has been given
only for the $\phi^4$ theory with one coupling
\cite{Eckmann75}. The field-theoretical RG series for models with
several couplings were analyzed {\it as if} they are asymptotically
divergent without a proof of this property. Moreover, there exists
strong evidence of possible Borel non-summability of the series
obtained for disordered models \cite{Griffits69}.

Recently, it was demonstrated analytically, that the RG functions of
the $d=0$ random Ising model are Borel summable, if a special
resummation technique is exploited \cite{Alvarez00}, the application
of this technique to the $d=3$ RG series of the random Ising model
allows to restore the convergence of the results obtained in the
massive RG scheme \cite{Pelissetto00} and in the minimal subtraction
scheme \cite{Blavatska01}.

In our case, the question about the summability of the series
(\ref{beta})-(\ref{gamma}) remains open. Nevertheless, we apply
various kinds of resummation techniques to obtain reliable
quantitative results. In the following section, we describe the procedure
of the Chisholm-Borel resummation technique, which proves to be most
effective in our case, and we present numerical results for the critical
behavior of systems with extended defects both in the 3-dimensional case
and in the case of general (non-integer) dimension.


\section{The results}


\subsection{Resummation procedure}

The two-variable Chisholm-Borel resummation technique
\cite{Jug83,reviews} consists of several steps. To explain, how it is
applied to the RG functions let us denote any of
the expressions (\ref{beta})--(\ref{gamma}) by
$f\equiv f(u,v)$.  First, we construct the Borel image of the initial
RG function $f$:
$$
f=\sum_{i,j}a_{i,j}u^iv^j\to\sum_{i,j}\frac{a_{i,j}(ut)^i(v
t)^j}{\Gamma(i+j+1)},
$$
where $\Gamma(x)$ is Euler's gamma
function.  Then, the Borel image is extrapolated by
a rational Chisholm \cite{Chisholm73} approximant $[K/L](u,v)$.
One constructs this ratio of two polynomials
of order  $K$ and $L$
such that its truncated Taylor
expansion is equal to that of the Borel image of
the function $f$. The resummed function is then calculated by an
inverse Borel transform of this approximant:
\begin{equation}\label{res}
f^{res}=\int_0^{\infty}{\rm d}t \exp(-t)[K/L](ut,vt).
\end{equation}
There is a lot of possibilities to choose a Chisholm approximant in
two variables.  The most natural way is to construct it such that, if
any of $u$ or $v$ is equal to zero, the familiar results are obtained
for reduced model. Here, for the Borel images of the $\beta$-functions we
have chosen the following approximant with a linear denominator:
\begin{equation}
\beta^{chis}=\frac{b_{0,0}+b_{1,0}ut+b_{0,1}vt+b_{1,1}uvt^2}{1+
c_{1,0}ut+c_{0,1}vt}.
\label{aprox}
\end{equation}
Note, that the polynomial in the numerator is chosen to be symmetric
in the variables $u$ and $v$. As far as the Chisholm approximants
enter the integrals (\ref{res}), the expression under integration may
contain poles; in this case an analytic continuation is needed,
taking the principal values of the integral.  Below we display results
only for those situations that do not require such an analytic
continuation.

To calculate the values of the critical exponents
$\nu_{\perp}$,$\nu_{||}$, $\gamma$ we substitute (\ref{gamma}) into
(\ref{exp}), apply the resummation procedure to the resulting
series, and, finally, substitute the obtained values of stable fixed
point coordinates in the above expressions for the critical
exponents.  Note, that while we use  the Chisholm approximant in the form
(\ref{aprox}) to analyze the series for
$\nu_{\perp}$,$\gamma$, the symmetry of the series for $\nu_{||}$
allows us  to use the  Chisholm approximant in the form:
\begin{equation}
\nu_{||}^{chis}=\frac{b_{0,0}+b_{1,0}ut+b_{0,1}vt}{1+
c_{1,0}ut+c_{0,1}vt}.
\label{aprox1}
\end{equation}

In the following we apply  the above resummation technique to analyze the
critical behavior of the model (\ref{hamiltonian}) both at space
dimensionality $d=3$ and in general dimensions.

\subsection{3-dimensional systems with extended impurities}

Let us first consider the model Eq. (\ref{hamiltonian}) for the
three-dimensional case.  Fixing the value $\varepsilon=1$ (i.e. $d=3$)
and treating $\varepsilon_d$ as a variable parameter, we look for the
common zeros of the resummed functions ${\beta_u}/{u}$ and
${\beta_{v}}/{v}$.  Let us note that there are two possible scenarios
for the change of FP stability under disorder introduced to the
system. This is due to the fact that the universality
class of the weakly diluted $m$-vector model
with point-like defects ($\varepsilon_d=0$)
depends in a crucial way on the order parameter
dimensionality $m$. There exists a marginal value $m_c$ such that for
$m>m_c$ the critical exponents of the $m$-vector model
remain unchanged by point-like defects
whereas for $m<m_c$ they split to new values.
Discussing the Harris criterion
we mentioned, that for integer $m$
point-like defects lead to a new universality class only for the
Ising model ($m=1$).  The present estimates for $m_c$ based on the
resummation of six-loop RG functions definitely imply $m_c<2$:
$m_c=1.942\pm0.026$ \cite{Bervillier86} and $m_c=1.912\pm 0.004$
\cite{Dudka01}.

\begin{figure} [htb]
\begin{center}
\includegraphics[height=42mm]{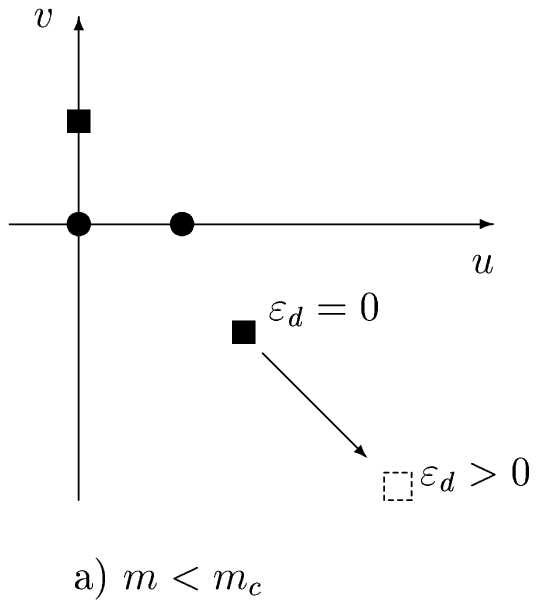}%
\includegraphics[height=42mm]{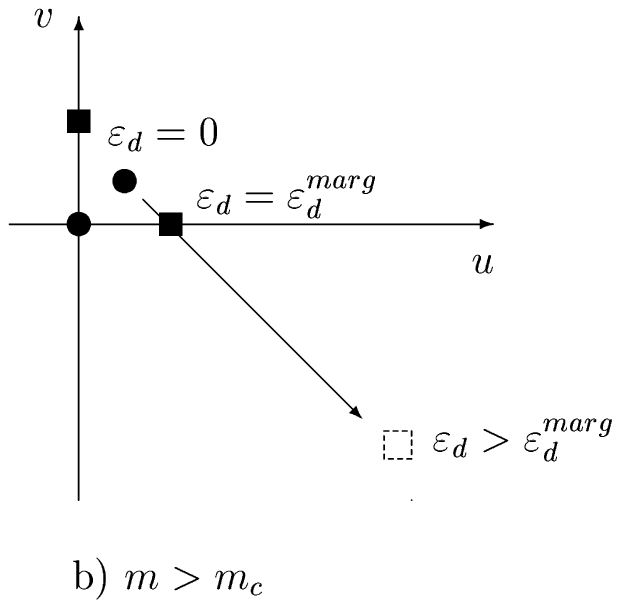}
\end{center}
\caption{\label{fig2}The two qualitative scenarios for the
  change of fixed point
  stability. Circles mark unstable fixed points, squares are the
  stable ones. The polymer FP ($u^*=0,v^*\neq0$) is stable but
  unphysical.  a) The spin dimensionality of the magnetic system is
  below the marginal one.  The random FP ($u^*\neq 0, v^*\neq 0$) is
  stable at $\varepsilon_d=0$.  As the parameter $\varepsilon_d$
  increases, the mixed fixed point remains stable and moves within the
  physically accessible region. b) The spin dimensionality of the
  magnetic system is above the marginal one. The pure FP ($u^*\neq
  0,v^*= 0$) is stable at $\varepsilon_d=0$, whereas the random FP is
  unstable and lies in the unphysical region. As $\varepsilon_d$
  increases, the mixed fixed point moves towards the physical
  region and at $\varepsilon_d=\varepsilon_d^{{{\rm marg}}}$
  becomes stable.}
  \end{figure}

For $m<m_c$, at $\varepsilon_d=0$ the random FP ($u\neq 0,v\neq 0$) is
stable and physically accessible. Any small increase of  the parameter
$\varepsilon_d$ leads to a shift of the stable FP value and therefore
to new critical exponents, compared to the pure ones (see Fig.
\ref{fig2}, a).  For $m>m_c$ the situation is different. Here,
the point-like ($\varepsilon_d=0$) disorder  is not relevant, and the
pure fixed point ($u^*\neq 0,v^*=0$) is stable, whereas
the random fixed point lies in the unphysical region and is unstable.
Switching on the parameter $\varepsilon_d$ leads to a shift
of the random FP, and at some value of this parameter the
random FP passes to the physically accessible  region, at the same
time it interchanges stability with the pure FP, i.e. it becomes stable (see
Fig. \ref{fig2}, b). This corresponds to a crossover to another
universality class with different critical exponents.

The the stable FP coordinates  obtained for the
3-dimensional $m$-component magnetic systems with $m=1,2,3,4$ are
presented in Table \ref{coord}.  One sees that for $m=2,3,4$
(i.e. for $m>m_c$) the pure FP ($u^*\neq0, v^*=0$) looses its
stability for some marginal value  $\varepsilon_d^{{\rm marg}}$.
As mentioned in the introduction, $\varepsilon_d^{{\rm marg}}$ can be
calculated from the correlation length
critical exponent of the pure magnet with help of the generalized Harris
criterion (\ref{criterion}). We will return to these estimates below.

The results obtained for the resummed values of the critical
exponents are  presented in Table \ref{expon}.
We may draw some conclusions from these results. As said above,
the case $\varepsilon_d=0$ describes point-like quenched disorder, so it
reproduces the well known results \cite{reviews}; in this case there
are no longitudinal components of the critical exponents.
For $\varepsilon_d\neq 0$,
the relation $\nu_{||}>\nu_{\perp}$ holds for every $\varepsilon_d$
and $m$. This may be explained in the following way:
the extended defects cut interacting paths of spins
perpendicular to the extended-defect direction, so in the parallel
direction the fluctuations are stronger and the correlation length
more sharply diverges.

\begin{figure}[htbp]
\begin{center}
{\includegraphics[width=85mm]{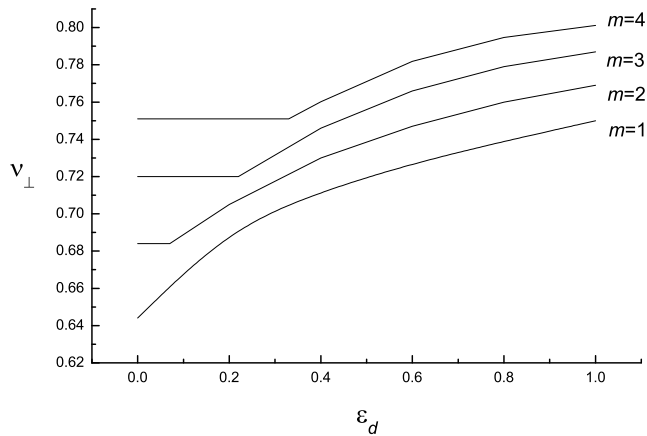}}
\end{center}
\caption{The critical exponent of the correlation length perpendicular to the
  $\epsilon_d$-dimensional ``lines" of extended defects as function of
  the parameter $\varepsilon_d$ for three-dimensional $m$-component
  spin systems.}
\label{fig3}
\end{figure}

\begin{figure}[htbp]
\begin{center}
{\includegraphics[width=85mm]{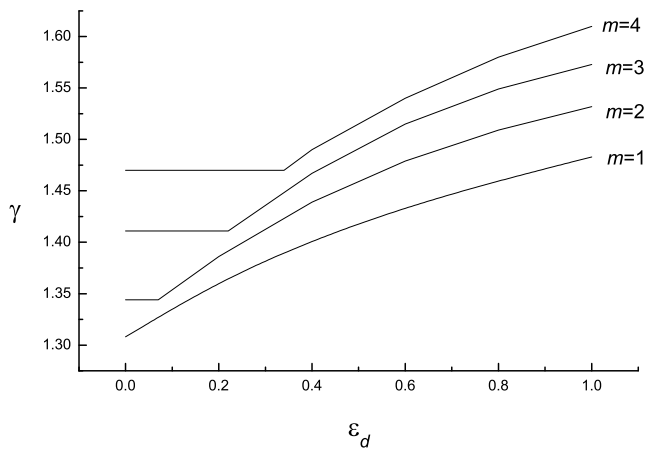}}
\end{center}
\caption{The critical exponent of the magnetic
  susceptibility of three-dimensional $m$-component spin systems as
  function of the parameter $\varepsilon_d$.}
\label{fig4}
\end{figure}

In Figs. \ref{fig3}, \ref{fig4} we plot the values of the critical
exponents $\nu_{\perp}$ and $\gamma$ for $m$-component systems with
$m=1,2,3,4$ as functions of the parameter $\varepsilon_d$.
They clearly represent our discussion above: as we can see,
in the case of the Ising model ($m=1$), any small increase of
$\varepsilon_d$ results in  a change of the critical exponents,
whereas for $m=2,3,4$ the critical exponents remain constant and
equal to the corresponding exponents of pure model, until
$\varepsilon_d$ reaches its marginal value, and only for
$\varepsilon_d>\varepsilon_d^{{{\rm marg}}} $ the values of
exponents increase.  Note, that as the parameter
$\varepsilon_d$ changes from 0 to 1, the resulting change in the
values of the critical exponents are sufficiently large to be experimentally
observed (e.g., the value of $\nu_{\perp} $ for Ising model
increases by 13 \%).  Thus, the influence of extended defects
on the critical behavior of three-dimensional magnetic systems can be
observed experimentally.

Let us return to the estimates of the marginal value
$\varepsilon_d^{{{\rm marg}}}$, at which the crossover to a new
universality class occurs. In Fig. \ref{fig1} we plot the
estimates for $\varepsilon_d^{{{\rm marg}}}$ as obtained from
six-loop results \cite{Guida98} for the correlation length
critical exponent $\nu_p(m)$ of the pure $m$-vector magnet:
$\varepsilon_d^{{{\rm marg}}}(m=1) =-0.173$;
$\varepsilon_d^{{{\rm marg}}}(m=2)=0.016$; $\varepsilon_d^{{{\rm
marg}}}(m=3) =0.172$; $\varepsilon_d^{{{\rm marg}}}(m=4) = 0.300$.
 The RG
functions of the $m$-vector magnet with extended impurities
exploited in this paper (expressions (\ref{beta})-(\ref{gamma}))
are currently known with two-loop accuracy. In this approximation
we get for the marginal value $\varepsilon_d^{{{\rm marg}}}(m=1)
=-0.105$; $\varepsilon_d^{{{\rm marg}}}(m=2)=0.077$;
$\varepsilon_d^{{{\rm marg}}}(m=3) =0.222$; $\varepsilon_d^{{{\rm
marg}}}(m=4) = 0.337$. The phase boundary of the critical behavior
of 3-dimensional $m$-component magnetic systems in the
$m,\varepsilon_d$-plane is shown in Fig. \ref{fig1} by the dashed
curve for the  two-loop approximation. It is remarkable that the
resummed two-loop and six-loop results for $\varepsilon_d^{{{\rm
marg}}}$  lead to a very similar phase diagram. In contrast to
this the first  order $\varepsilon, \varepsilon_d$-expansion
predicts a qualitatively different behavior \cite{Boyanovsky82}:
the random FP is in the physical region for
$\varepsilon>\varepsilon_d$, and randomness is relevant for
$m<m_c=4(\varepsilon+
2\varepsilon_d)/(\varepsilon-\varepsilon_d)$, while for $m>m_c$
the random FP lies in the unphysical region and the pure FP is
stable, so for $\varepsilon>\varepsilon_d$ the stability changes
at $m_c$. For $\varepsilon<\varepsilon_d$ the only physical fixed
point is random and is always stable.  From these first order
results one would conclude, that the disorder is relevant for
every $m<4$ and positive $\varepsilon_d$.

Another interesting question concerns the existence of an upper
marginal value for the defect dimensionality $\varepsilon_d$. This
question has not been raised in previous works, where the double
$\varepsilon, \varepsilon_d$-expansion has been exploited. In our
analysis, we observe the disappearance of a stable reachable FP
for $\varepsilon_d$ slightly above $1$. This, in principle,
may serve as evidence of an upper boundary for $\varepsilon_d$
above which the second order phase transition ceases to exist.
However, this behavior was observed in the region of couplings
$u,v$ where the Chisholm approximants (\ref{aprox}) that enter the
expressions for the resummed $\beta$-functions (\ref{res}) have
poles on the positive real axis. This excludes a definite answer
about the presence and stability of the fixed points for high
$\varepsilon_d$. On the other hand, simple reasons support the
idea of an upper boundary for $\varepsilon_d$. Interpreting
$\varepsilon_d$ as the fractal dimensionality of the defects, it
is clear, that it cannot exceed the dimensionality of the
embedding space, $d=3$. Furthermore, one expects physically, that
extended defects of large dimension (e.g. parallel planar defects
with $\varepsilon_d=2$), that extend throughout the system, divide
the system into non-interacting regions and thus inhibit
ferromagnetic order.

\subsection{Systems with extended defects in general dimensions}

Let us now consider the case of non-integer space dimension $d$. The
concept of non-integer space dimension is common in the theory of
critical phenomena.  Considering $d$ as a continuous variable is not
purely formal; for the critical behavior of spin systems placed on the
sites of self-similar fractal lattices \cite{Mandelbrot} it has a
direct geometrical interpretation.

Considerable effort has been made to calculate the critical exponents
for the second order phase transition for general (non-integer) values
of $d$ (see Refs. \cite{LeGuillou87,Holovatch93,Holovatch97} for
review). Most papers study the Ising model. The critical exponents of
the pure $m$-vector model in general dimensions have been calculated
in the field-theoretical RG scheme for the case $m=1$ (Ising-like
systems) \cite{LeGuillou87,Holovatch93} and for $m=2$, $3$,
$4$\cite{Holovatch97}. Numerical results for the weakly
diluted Ising model are presented in Ref. \cite{dilHol}.

To analyze the critical properties of the model (\ref{hamiltonian})
for general dimensions, the 3d approach, used above for the RG
functions of the present model is generalized. We treat the RG
functions directly at different fixed values of (non-integer) space
dimension $d$ and defect dimensionality $\varepsilon_d$.

Again we are looking for the marginal value
of the parameter $\varepsilon_d$ for models at different non-integer
space dimensions.  To make use of the generalized Harris criterion
(\ref{criterion}) in
this case, we take the known estimates from a 3-loop RG analysis of the
critical exponent $\nu_p(d)$ for the pure $m$-vector model at $m=1$
\cite{Holovatch93} and $m=2,3,4$, \cite{Holovatch97}.
In accordance with the Mermin-Wagner-Hohenberg theorem \cite{Mermin66},
the lower critical dimension of the $m$-vector model is $d_L=2$ for
$m\geq 2$, while in the case of the Ising model ($m=1$) one has
$d_L=1$ \cite{Stanley71}  (for $d\leq d_L$,  there is no spontaneous
magnetic ordering at $T>0$). In order to take into account these conditions,
and also to reproduce the exact Onsager value \cite{Onsager44} for the 2d
Ising model ($\nu=1$), a more sophisticated resummation procedure was
exploited in Refs. \cite{Holovatch93,Holovatch97}, and we will use these
results.  \begin{figure}[htbp]
\begin{center}
{\includegraphics[width=85mm]{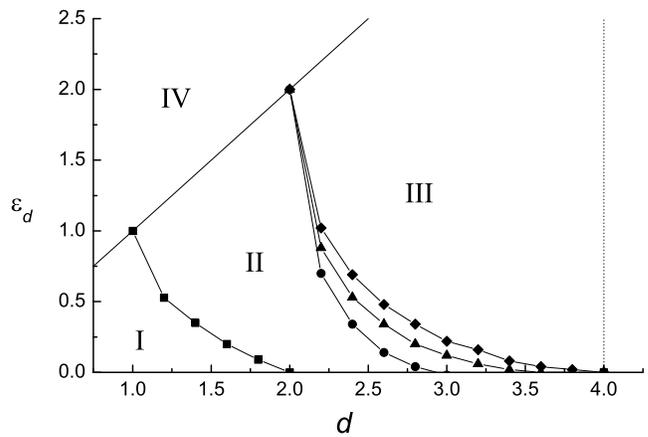}}
\end{center}
\caption{The phase diagram of  $m$-component
  magnetic systems with extended defects in the
  $d,\varepsilon_d$-plane.  The lower marginal value of
  $\varepsilon_d$ for $m=1$ (squares), 2 (circles), 3 (triangles), 4
  (diamonds) is shown. In region I disorder with extended defects is
  irrelevant for all $m\geq1$; region II: disorder is relevant only
  for the Ising model ($m=1$); region III: disorder with extended
  defects is relevant for systems with $m=2, 3, 4$.  The solid
  diagonal line corresponds to $\varepsilon_d=d$: the region IV
  corresponds to non-physical situations.} \label{fig5}
\end{figure}

Substituting the known data \cite{Holovatch93,Holovatch97} for
$\nu(d)$ into the generalized Harris criterion (\ref{criterion}) we
find the marginal value of the parameter $\varepsilon_d$ for different
$m$-component systems as function of space dimension, see
Fig. \ref{fig5}. This figure serves as a phase diagram in parameter
space ($d$, $\varepsilon_d$). It should be noted, that, as follows
directly from the Harris criterion (\ref{harris}), one has
$\varepsilon_d^{{{\rm marg}}}=0$ at $d=4$ (the upper critical
dimension, where the Gaussian FP ($u^*=0,v^*=0$) is stable and
$\nu=1/2$), whereas at $d=d_L$, one has $\varepsilon_d^{{{\rm
      marg}}}=d_L$, independent of $m$.  Note, that for $m=1$, $2$
there exists a range of space dimensions $d$, at which
$\varepsilon^{\rm marg}_d<0$, and, thus, the presence of any extended defects
causes changes to the critical behavior. In the region marked as I
in Fig. \ref{fig5} disorder with extended defects is irrelevant for
all the systems under consideration; in region II it becomes relevant
for the Ising model, passing from the region II to III it gradually
becomes relevant for the models with $m=2, 3, 4$.  As discussed above,
the concept of an upper marginal value for the parameter
$\varepsilon_d$ naturally arises. The defect dimensionality
$\varepsilon_d$ cannot exceed the dimension of the embedding space
$d$, and thus cannot take values from the region marked as IV in Fig.
\ref{fig5}.

To get the values of the critical exponents of the set of systems
with extended defects at general dimensions $d$, we exploit the same
2-variable Chisholm-Borel resummation technique as for the
3-dimensional case above. But now we calculate the RG functions
(\ref{beta})-(\ref{gamma}) for fixed non-integer $\varepsilon=4-d$.
We present here only the results obtained for the extended-defect
Ising model ($m=1$), as far as it is most sensitive to
introducing defects in a wide region of space dimensionality $d$,
as can be observed from Fig. \ref{fig5}.

\begin{figure}[htbp]
\begin{center}
{\includegraphics[width=85mm]{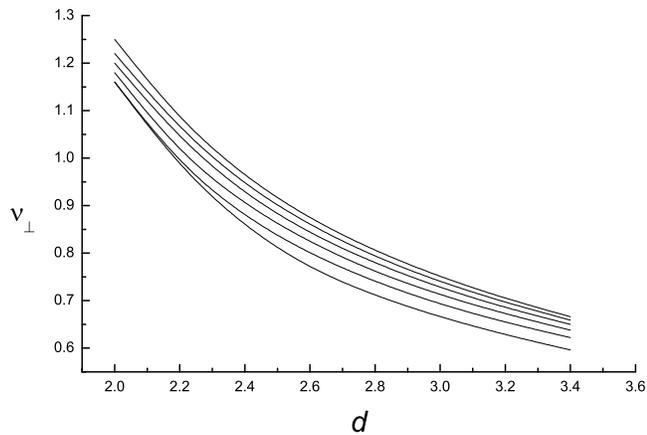}}
\end{center}
\caption{The critical exponent of the correlation length
  perpendicular to the $\varepsilon_d$-dimensional ``lines" of
  extended defects, as function of dimension $d$ for Ising
  systems at values of the parameter $\varepsilon_d=0,0.2, 0.4, 0.6,
  0.8, 1$ from below.}
\label{fig6}
\end{figure}
\begin{figure}[htbp]
\begin{center}
{\includegraphics[width=85mm]{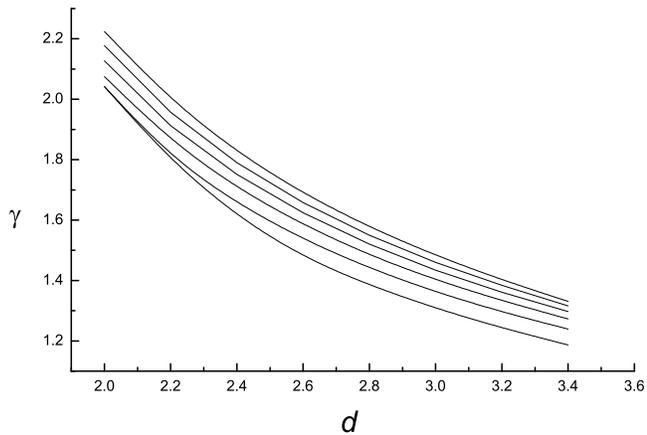}}
\end{center}
\caption{The critical exponent of the magnetic susceptibility
  as a function of the dimension $d$ for Ising systems at values of
  the parameter $\varepsilon_d=0, 0.2, 0.4, 0.6, 0.8, 1$ from below.}
\label{fig7}
\end{figure}

Before discussing the results obtained
for general $d$ let us note, that their accuracy depends on the
distance $\varepsilon=4-d$ from the upper critical dimension $d=4$
\cite{LeGuillou87,Holovatch93,Holovatch97,dilHol};
for larger expansion parameters  $\varepsilon$
also the fixed point values of the coupling
constants grow leading to expansions which are
difficult to estimate, even when resummation procedures are applied.
We demonstrate this for the $d=2$ and $d=3$ Ising model
with point-like quenched disorder ($\varepsilon_d=0$).
The Onsager solution of the $d=2$ pure Ising model
\cite{Onsager44} brings about a logarithmic divergence of the
specific heat, i.e. an exponent $\alpha_p=0$.
With the Harris criterion
(\ref{harris}) the well known irrelevance of
this type of disorder for the $d=2$ Ising model follows. The correlation
length critical exponent in this case equals \cite{Onsager44}
$\nu_p=1$, leading to $\varepsilon_d^{{\rm marg}}=0$ by applying
the generalized Harris criterion (\ref{genharris}). However, in
our 2-loop calculations refined by the Chisholm-Borel resummation
technique we find  for the $d=2$ Ising model correlation length
and magnetic susceptibility critical exponents
$\nu_p=1.162$ and  $\gamma_p=2.042$ correspondingly. These
values  differ by 16 \% from their exact values \cite{Onsager44}
$\nu=1$, $\gamma_p=7/4$. This implies
a marginal value of the
extended defects dimensionality of
$\varepsilon_d^{{\rm marg}}=0.278$.
For $d=3$, we reproduce the two-loop results of the random Ising
model critical exponents, obtained in the minimal
subtraction scheme in Ref. \cite{Folk00} $\nu_{\perp}=0.665$,
$\gamma=1.308$. They differ from the most accurate six-loop
estimates \cite{Pelissetto00} $\nu_{\perp}=0.678(10)$,
$\gamma=1.330(17)$ by only of 2 \%. Again, for the marginal value of the defect
dimensionality  we find $\varepsilon_d^{{\rm
marg}}=-0.105$, the six-loop value being $\varepsilon_d^{{\rm
marg}}=-0.173$ (see Fig. (1)). To conclude, we expect the accuracy
of our results for the critical exponents to be of the order of
several percents for $d=3$ and to increase to the order of $\sim 10\%-20\%$
for $d=2$, (recall that the approach
gives the exact data for $d=4$). For this reason we do not show
any results for $d<2$, taking that the confidence interval
for them is even larger. On the other hand,
the two-loop approximation restricts our investigation to dimensions
$d$ smaller than $3.4$; namely, for $d\geq 3.4$ the rational
approximants that enter the integrals for the resummed functions have
poles that exclude reliable calculations in this case.

The values of the critical exponents $\nu_{\perp}$ and $\gamma$
are plotted as functions of $d$ parameterized by $\varepsilon_d$ in Figs.
\ref{fig6}, \ref{fig7} .
The parameter $\varepsilon_d$ of the different curves changes from 0 to 1 by steps of 0.2.
For values of $d$ below $2.1$ the critical exponents remain
equal to those of the pure model  for small $\varepsilon_d$, until $\varepsilon_d$
reaches its marginal value (of about $\varepsilon_d\simeq 0.27$
in the present two-loop calculations); then, a crossover to the
diluted regime takes place, and the exponents increase
continuously with increasing the parameter $\varepsilon_d$ at fixed
space dimension $d$.
We interpret this by noting
that
an increase of  $\varepsilon_d$ leads to an effective
decrease of the space dimension implying a sharper
divergency of the correlation  length
(characterized by a larger value of $\nu$).
The results for $\nu_{||}$, not shown here,
display the same tendencies under changes of the parameters
$d$, $\varepsilon_d$. Let us note however, that the relation
$\nu_{||}>\nu_{\perp}$ holds for all $d$, $\varepsilon_d$, as
observed for $d=3$ and explained in the previous subsection.


\section{Conclusions}

In our study we provide numerical estimates for the critical
exponents of $m$-component magnetic systems in the
presence of extended macroscopic defects.  The impurities
are envisaged as $\varepsilon_d$-dimensional objects, extending
throughout the system, whereas in the remaining $d-\varepsilon_d$
dimensions  they are randomly distributed.  The presence of these
impurity ``lines'' introduces anisotropy into the systems; as a
result, the parallel and transverse correlation lengths naturally
arise and are described in the vicinity of the critical point by the
respective correlation length  critical exponents
\cite{Dorogovtsev80}.  Such systems are not covered by the original
Harris criterion \cite{Boyanovsky82};  disorder with extended
impurities is relevant over a wider range of $m$ and $d$ than the
point defect disorder  (see Figs. \ref{fig1}, \ref{fig5}).

Although the model of $m$-component magnetic system with extended
quenched defects has attracted much attention and serves as a subject
of study in a number of works, most results are based on the double
expansion in parameters $\varepsilon,\varepsilon_d$ and are rather of
a qualitative character.  We analyze the RG functions of the model,
obtained in the minimal subtraction
scheme\cite{Boyanovsky82,Lawrie84}, treating them directly for fixed
$d$ and fixed parameter $\varepsilon_d$.  To evaluate the RG
functions we apply appropriate resummation techniques, that have
proven to be fruitful in the analysis of point-like quenched disorder
\cite{reviews}. In our case, the question about the summability of the
corresponding RG functions is open; nevertheless, in order to obtain
reliable quantitative results, we apply a simple two-variable
Chisholm-Borel resummation technique.  The values of the stable fixed
point coordinates and critical exponents for 3-dimensional systems
are extracted, as well as the estimate for the lower marginal value
for parameter $\varepsilon_d$ is obtained. The case $\varepsilon_d=0$
describes point-like quenched disorder and reproduces well-known
results \cite{reviews}. For $\varepsilon_d>0$ it was found, that the
relation $\nu_{||}>\nu_{\perp}$ holds for every $\varepsilon_d$ and
$m$, which describes the following physical situation: the extended
defects cut interacting paths of spins perpendicular to the
extended-defect direction, so in the parallel direction the
fluctuations are stronger and the correlation length more sharply
diverges.  The only numerical results for the exponents of the model
(\ref{hamiltonian}) were obtained so far for $\varepsilon_d=1$ using
the Pad\'e-analysis \cite{Lawrie84}, they are shown in the last line
of Table \ref{expon}.  We note however, that our present resummation
technique has shown its efficiency and accuracy in studies of models
with point-like structural disorder \cite{reviews} and is generally
known to provide more reliable estimates as compared to the simple
Pad\'e-analysis.

 The critical behavior of the
3-dimensional Ising model ($m=1$) is changed by an extended
impurity characterized by any $\varepsilon_d\neq0$,
 whereas for $m=2,3,4$ the extended defect
disorder is relevant only for $\varepsilon_d>\varepsilon_d^{{{\rm marg}}}$.
We estimate the marginal value of the parameter $\varepsilon_d$ also
for magnetic systems in general (non-integer) space dimensions. In
Figs. \ref{fig6}--\ref{fig7} we plot the results for the Ising
model in general dimensions, which appears to be the most
``sensitive'' to introducing extended defects in a rather wide
range of space dimensionality $d$.

The concept of an upper marginal value of  $\varepsilon_d$
naturally arises in studies of magnetic systems with extended
impurities.  Interpreting $\varepsilon_d$ as a fractal dimensionality
of defects, it is clear, that  it cannot exceed the dimensionality of
embedding space, $d$. Moreover, physically one may expect, that
extended defects of large dimension (e.g. parallel planar defects with
$\varepsilon_d=2$), each extending throughout the system, will divide
the system into non-interacting regions and thus prevent it from
ferromagnetic ordering.

Unfortunately, there is no simulational investigation of
systems with parallel extended defects that we know of \cite{note}.
However, our calculations bring about an essential change in
critical exponents due to presence of extended impurities (e.g.
$\nu_{\perp}$ for the Ising model with extended impurities
increases by 13 \% as the parameter $\varepsilon_d$ changes from 0
to 1). We hope that the influence of this kind of disorder on
the critical behavior of three-dimensional magnetic systems may
create some interest for numerical simulations or experimental
measurements.

\section*{Acknowledgements}

V. B. gratefully acknowledges support by DAAD (Deutscher Akademischer
Austausch Dienst) and hospitality of the Theoretical Polymer
Physics group at Freiburg University.


\begin{table}
\begin{center}
\begin{tabular}{|c| c c| c c| c c| c c |}
\hline
&  $m=1$  &  &  $m=2$ &
& $m=3$&  &$m=4$ & \\
\hline
$\varepsilon_d$  &  ${u}^*$&  ${v}^*$&  ${u}^*$&
${v}^*$&  ${u}^*$&  ${v}^*$ &  ${u}^*$&  ${v}^*$\\
\hline
0& 1.5772& -0.2416& 1.1415& 0 & 1.0016& 0 & 0.8877 & 0\\
0.1& 1.7640& -0.4187 & 1.1688& -0.0372 & 1.0016& 0 & 0.8877& 0 \\
0.2& 1.9169& -0.5635 & 1.2713& -0.1857  & 1.0016& 0 &0.8877&0\\
0.3& 2.0478& -0.6859 & 1.3509& -0.3117& 1.0467& -0.1028&0.8877&0\\
0.4& 2.1633& -0.7919 & 1.4145& -0.4195 & 1.0908& -0.2186 &0.9069 & -0.0771\\
0.5& 2.2671& -0.8853 & 1.4665& -0.5125& 1.1238& -0.3188&0.9290 & -0.1869\\
0.6& 2.3619& -0.9688 & 1.5096& -0.5932 & 1.1486& -0.4053&0.94346 & -0.2819\\
0.7& 2.4493& -1.0445 & 1.5457& -0.6635 & 1.1672& -0.4799&0.9524 & -0.3635\\
0.8& 2.5306& -1.1131 & 1.5763& -0.7249 & 1.1812& -0.5440&0.9575 & -0.4331\\
0.9& 2.6067& -1.1762 & 1.6025& -0.7788  & 1.1917& -0.5991&0.9598 & -0.4919\\
1.0& 2.6783& -1.2343 & 1.6249& -0.8261 & 1.1995& -0.6461&0.9602&-0.5412\\
1.1& 2.7466& -1.2896 & 1.6443& -0.8675& 1.2053& -0.6858&0.9594&-0.5819\\
\hline\end{tabular}\end{center}
\caption{The coordinates of the stable fixed points
of the RG equations
for 3-dimensional magnetic systems with extended
$\varepsilon_d$-dimensional impurities for different $m$. Note, that
the pure fixed point ($u^*\neq 0,v^*= 0$) looses its stability for
some marginal value $\varepsilon_d^{{{\rm marg}}}(m)$.}
\label{coord}
\end{table}
\begin{table}
\begin{center}
\begin{tabular}{|c| c c c|  c c c| c c c| c c c|}
\hline &   &  $m=1$  &  & & $m=2$ & & &$m=3$& & & $m=4$& \\\hline
$\varepsilon_d$  &
$\nu_{||}$  &  $\nu_{\perp} $  & $\gamma$ &
$\nu_{||}$  &  $\nu_{\perp} $  & $\gamma$&
$\nu_{||}$  &  $\nu_{\perp} $  & $\gamma$&
$\nu_{||}$  &  $\nu_{\perp} $  & $\gamma$
\\
\hline
0  & -- & 0.665& 1.308 & -- & 0.684& 1.344& -- & 0.720& 1.411& -- &0.751&1.470 \\
0.1 & 0.714& 0.680& 1.338& 0.691& 0.688& 1.352& 0.720& 0.720& 1.411&0.751&0.751&1.470\\
0.2 & 0.741& 0.692& 1.362 & 0.719& 0.705& 1.386 & 0.720& 0.720& 1.411&0.751&0.751&1.470\\
0.3& 0.765& 0.702& 1.384  & 0.744& 0.718& 1.414 & 0.740& 0.732& 1.438&0.751&0.751&1.470\\
0.4 & 0.786& 0.712& 1.402& 0.766& 0.730& 1.439 & 0.763& 0.746& 1.467&0.761&0.760& 1.491\\
0.5 & 0.805& 0.720& 1.419  & 0.785& 0.739& 1.460 & 0.784& 0.757& 1.493&0.784&0.772&1.520\\
0.6 & 0.822& 0.727& 1.434 & 0.802& 0.747& 1.479 & 0.801& 0.766& 1.515&0.811&0.782&1.545 \\
0.7& 0.838& 0.733& 1.448 & 0.818& 0.754& 1.495 & 0.817& 0.773& 1.533&0.837&0.789&1.565 \\
0.8 & 0.853& 0.739& 1.460 & 0.831& 0.760& 1.509 & 0.831& 0.779& 1.549&0.858& 0.795&1.582 \\
0.9 & 0.867& 0.745& 1.472 & 0.843& 0.765& 1.522 & 0.842& 0.783& 1.562&0.875&0.799&1.595\\
1.0 & 0.880& 0.750& 1.483 & 0.854& 0.769& 1.532  & 0.852& 0.787& 1.573&0.886&0.801&1.605\\
1.1 & 0.892& 0.754& 1.493 & 0.863& 0.773& 1.542  & 0.860& 0.789& 1.581&0.895&0.803&1.613\\
\hline
1.0 \cite{Lawrie84}& 0.84 & 0.67 & 1.34  & 0.60 & 0.56 & 1.13   & 0.66 & 0.61 & 1.24 & & &\\
\hline
\end{tabular}\end{center}
\caption{The critical exponents for 3-dimensional magnetic
systems with extended
$\varepsilon$-dimensional impurities for different $m$. For comparison
the last line shows the results of Ref. \protect \cite{Lawrie84}.}
\label{expon}
\end{table}

\end{document}